\documentclass[twocolumn,preprintnumbers,amsmath,amssymb,aps,prb,longbibliography]{revtex4-1}
\usepackage{graphicx}

\begin{document}

\title{Vortex Guidance and Transport in Channeled Pinning Arrays}
\author{C. Reichhardt and  C. J. O. Reichhardt}
\affiliation{Theoretical Division and Center for Nonlinear Studies,
  Los Alamos National Laboratory, Los Alamos, New Mexico 87545, USA
}

\date{\today}

\begin{abstract}
We numerically examine vortices interacting with pinning arrays where a portion of the pinning sites have been removed in order to create coexisting regions of strong and weak pinning.  The region without pinning sites acts as an easy-flow channel.  For driving in different directions with respect to the channel, we observe distinct types of vortex flow.  When the drive is parallel to the channel, the flow first occurs in the pin free region followed by a secondary depinning transition in the pinned region.  At high vortex densities there is also an intermediate plastic flow phase due to the coupling between the weak and strong pinning regions.  For driving applied perpendicular to the channel, we observe a jammed phase in which vortices accumulate on the boundary of the pinned region due to the vortex-vortex repulsion, while at higher drives the vortices begin to flow through the pinning array.  For driving at an angle to the channel, depending on the filling we observe a drive-induced reentrant pinning effect as well as negative differential mobility which occurs when vortices move from the unpinned to the pinned portion of the sample.

\end{abstract}
\maketitle

\vskip2pc

\section{Introduction}
Depinning and sliding dynamics of an assembly of particles in the presence of
either a random or a periodic landscape occurs
in a wide variety of condensed matter systems,
including vortices in type-II superconductors \cite{Bhattacharya93,Olson98a},
sliding Wigner crystals \cite{Reichhardt01},
skyrmions \cite{Nagaosa13,Reichhardt15a},
colloids \cite{Bohlein12,McDermott13a}, 
and frictional dynamics \cite{Vanossi13}.
In most of these systems the disorder is generally uniform, but there are
some cases in which
the pinning is
anisotropic or spatially inhomogeneous,
where regions of strong pinning coexist
with regions of weak pinning.
An example of this is vortices in type-II superconducting samples that
have thicknesses modulations
\cite{Daldini74,Besseling05,Yu07,Shklovskij06,Shklovskij09,Yu10,Dobrovolskiy10,LeThien16,Dobrovolskiy17a,Dobrovolskiy17b,Dobrovolskiy17,Dobrovolskiy19}.
Here the pinning is anisotropic, and the vortices can slide easily in one direction
but not in the other.
When a driving current is applied such that the resulting force on the vortices is
aligned with the easy flow direction,
the pinning is stronger in thick regions than in thin regions,
and the vortices preferentially flow
in the weak pinning regions
past other vortices that are trapped in the strong pinning
regions \cite{Besseling05,Yu07,Dobrovolskiy17a}.
It is also possible to drive vortices at an angle with 
respect to a pinning array in order to
produce
a combination of guided flow and plastic vortex flow
\cite{Shklovskij06,Shklovskij09,Dobrovolskiy17b,Dobrovolskiy19}. 
Large scale regions of
strong pinning coexisting with weak pinning
can be created using masked irradiation techniques
\cite{Marchetti99,Basset01,Kwok02,Crassous11,Dobrovolskiy12,Trastoy14}. 
In soft matter and magnetic systems,
inhomogeneous pinning can appear when
one part of the sample contains strong disorder and the other
part of the sample  contains weak disorder \cite{Seshadri93,Reichhardt19c}. 
Previous work on inhomogeneous pinning has generally focused on the case when the
particles are driven parallel to an easy flow direction of the pinning,
which produces shear banding or gradients in the particle velocity.

Advances in nanostructuring have now made it possible to create carefully controlled
pinning landscapes.
For vortices in type-II superconductors,
there are numerous ways to generate
periodic pinning structures \cite{Baert95,Martin97,Reichhardt98a,Berdiyorov06,Swiecicki12},
so it should be feasible to create samples containing
strips of periodic pinning coexisting with
regions in which there are no
artificial pinning sites.
The vortex dynamics in such a system should
have several interesting features, since even 
in systems with uniform periodic pinning,
a rich variety of dynamic phases can arise
depending on the ratio of the number of vortices to 
the number of pinning sites,
the direction of the drive with respect to the pinning array symmetry,
and the size of the pining sites
\cite{Reichhardt97,Reichhardt99,Gutierrez09,Avci10,daSilva11}.

In this work we numerically examine a superconducting system
in which one half of the sample contains
a square pinning array and the other half
contains no pinning.
The initial vortex density in the absence of driving is uniform across the system.
Under an applied drive we
observe very different dynamical phases
depending on whether the driving direction is parallel, perpendicular, or at an angle
to the pinning stripe.
These dynamical phases are associated with
distinct features in the velocity-force curves and vortex flow patterns.
The
dynamics also depends strongly on the ratio of the number
of vortices to the number of pinning sites.
When an increasing drive is applied parallel to the pinning strip,
flow of only the vortices in the unpinned region is followed by
plastic flow at the interface between the pinned and unpinned regions and
then by a flow of all the vortices in the entire sample.
For perpendicular driving,
we observe a jammed phase
in which vortices accumulate 
along the edge of the pinned portion of the sample.
The most interesting regime is for driving at angle.  Here, the vortex motion is
confined to the pin-free channel and guided by the pinning strip for low drives,
but at sufficiently high drives,
moving vortices begin to enter the strong pinning region,
producing either
a reentrant pinned phase or negative differential mobility.
We map out these phases in a series of dynamic phase diagrams. 
Our results could be tested with nanostructured superconductors,
and should also be general to other assemblies of interacting particles
in inhomogeneous pinning.

\section{Simulation}

\begin{figure}
\includegraphics[width=3.5in]{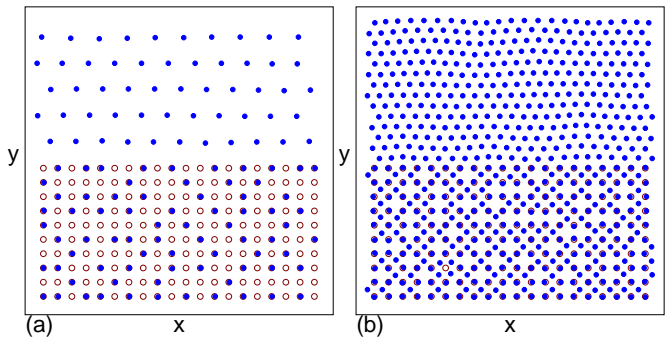}
\caption{
  The vortex positions (filled circles) and pinning site positions (open circles)
  in a system where half of the sample contains a square pinning array and the other
  half of the sample contains no pinning.
  (a) $B/B_{\phi} = 1/3$, where there are fewer
  vortices than pinning sites.
  (b) $B/B_{\phi} = 1.67$, where there are more vortices than pinning sites
  and interstitial vortices appear within the pinned region.  
}
\label{fig:1}
\end{figure}

We model the vortices as individual point particles in a two dimensional
system
of size $L \times L$
with periodic boundary conditions.  The lower half of the sample
contains
a square pinning array with lattice constant $a$,
while the upper half of the sample is pin-free.
The interface between the pinned and unpinned regions is along the $x$ direction.
There are $N_{p}$ pinning sites and $N_{v}$ vortices,
and the vortex density is given by $B=N_v/L^2$.
We define the
matching vortex density $B_{\phi}$ to be the density at which the number of
vortices would be equal to the number of pinning sites if the entire system were
filled with pinning.  Thus, in our case, at $B=B_{\phi}$, $N_v=2N_p$.
The initial vortex positions are obtained by
annealing from a high temperature down to $T = 0$,  which creates a uniform
vortex density with a roughly equal number of vortices in the pinned and pin-free
regions of the sample.
In Fig.~\ref{fig:1} we show a snapshot
of the vortex configurations after annealing for two different vortex densities.
At $B/B_{\phi} = 1/3$ in Fig.~\ref{fig:1}(a),
a portion of the pinning sites are empty, while 
at $B/B_{\phi} = 1.67$
in Fig.~\ref{fig:1}(b),
all of the pinning sites are occupied and a portion of the vortices
in the pinned region are located at interstitial positions between the pinning sites.  
In both cases, the vortices in the pin-free region form a hexagonal lattice.   

The vortex dynamics for vortex $i$ is obtained using the following equation of motion:
\begin{equation} 
\alpha_d {\bf v}_{i}  =
{\bf F}^{ss}_{i} +  {\bf F}^{p}_{i} + {\bf F}^{D}_{i} .
\end{equation}
Here the repulsive vortex-vortex interaction force is
${\bf F}_{i} = \sum^{N}_{j=1}K_{1}(r_{ij}){\hat {\bf r}_{ij}}$,
where
${\bf v}$ is the vortex velocity,
$\alpha_d = 1.0$ 
is the damping term,
$r_{ij} = |{\bf r}_{i} - {\bf r}_{j}|$ and
$K_{1}(r)$ is the modified Bessel function which falls
off exponentially for large $r$.
A uniform
driving force ${\bf F}^{D}=F_D{\bf \hat\alpha}$ is applied to
all the vortices, with
${\bf \hat \alpha}={\bf \hat x}$ for driving parallel to the pinning stripe,
${\bf \hat \alpha}={\bf \hat y}$ for driving perpendicular to the pinning stripe,
and
${\bf \hat \alpha}=\cos(\theta){\bf \hat x} + \sin(\theta){\bf \hat y}$
for driving at an angle $\theta$ to the pinning stripe.
The pinning sites are modeled as
finite range attractive parabolic potential wells
with a radius of $r_{p}$ which exert a maximum
pinning force of $F_{p}$ on a vortex.
In the single vortex limit, depinning occurs when $F_{D}/F_{p} > 1.0$.   
In this work we fix $B_{\phi} = 0.3$, $F_{p} = 1.0$, and $r_{p} = 0.35$,
and
we vary the number of vortices or filling
factor $N_v/N_p$ as well as the direction of the applied drive.
For each value of $F_D$, we measure the average velocity in the $x$ direction,
$\langle V_x\rangle=N_v^{-1}\sum_i^{N_v}{\bf v}_i\cdot {\bf \hat x}$,
or the $y$ direction,
$\langle V_y\rangle=N_v^{-1}\sum_i^{N_v}{\bf v}_i\cdot {\bf \hat y}$.

\section{Driving in the Parallel Direction}

\begin{figure}
\includegraphics[width=3.5in]{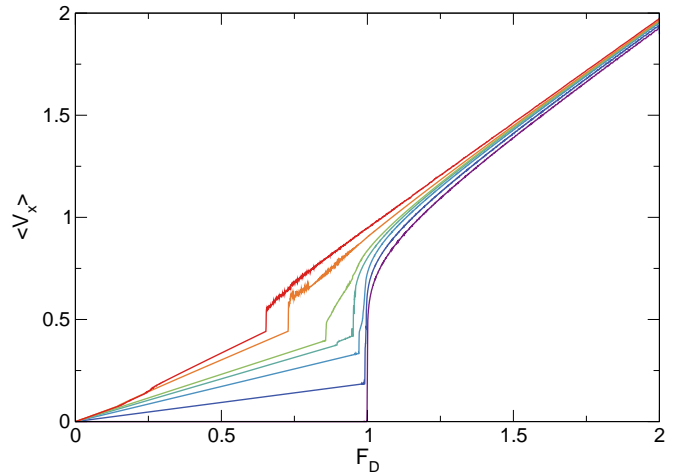}
\caption{$\langle V_{x}\rangle$ vs $F_{D}$ for the system in
  Fig.~\ref{fig:1} with parallel driving
  at $B/B_{\phi} = 0.083$, 0.167, 0.33, 0.5, 0.67, 1.33, and 1.67, from
bottom to top.
}
\label{fig:2}
\end{figure}

\begin{figure}
\includegraphics[width=3.5in]{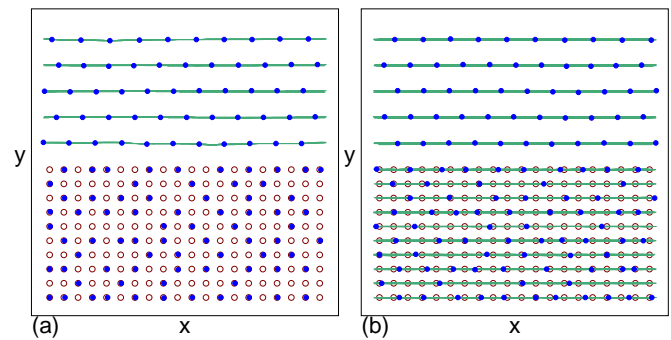}
\caption{
  The vortex positions (filled circles), pinning site positions (open circles),
  and vortex trajectories (lines) 
  for the system in Fig.~\ref{fig:2} with parallel driving at $B/B_{\phi}  = 0.33$.
  (a) At $F_{D} = 0.5$, only the vortices in the unpinned region
  flow.
  (b) At $F_{D} = 1.1$, all the vortices are flowing.
}
\label{fig:3}
\end{figure}

We first consider driving in the easy flow or $x$-direction.
In Fig.~\ref{fig:2} we plot
$\langle V_{x}\rangle$ versus $F_{D}$
for samples with $B/B_{\phi} = 0.08$, 0.167, 0.33, 0.5, 0.67, 1.33, and $1.67$.
When $0.08 < B/B_{\phi} < 0.33$,
we observe two stages of vortex flow.  At lower drives,
only the vortices in the unpinned region are moving,
while at higher drives, above the step in $\langle V_x\rangle$,
vortices in the pinned region depin and begin to move.
Interstitial vortices first begin to penetrate the edge of the pinned region
when $B/B_{\phi} > 0.33$.
The interstitial vortices are only indirectly pinned by interactions with the pinned
vortices, and they depin near
$F_{D} = 0.15$.  Their motion
produces
an intermediate flow state
for $B/B_{\phi}>0.33$
in which plastic or disordered flow of vortices occurs in the pinned portion of the sample
along with a more ordered flow of vortices in the unpinned portion of the sample.
For $ B/B_{\phi}  \leq 0.08$, all of
the vortices are pinned at low drives
and there is a single depinning threshold at $F_{D}/F_{p} = 1.0$.
In Fig.~\ref{fig:3}(a) we illustrate the vortex trajectories
for the system in Fig.~\ref{fig:2} with $B/B_{\phi} = 0.33$
at $F_{D} = 0.5$, where only vortices in the pinned region are flowing.
At $F_{D} = 1.1$ in the same sample,
Fig.~\ref{fig:3}(b) indicates that  all the vortices are now flowing.

\begin{figure}
\includegraphics[width=3.5in]{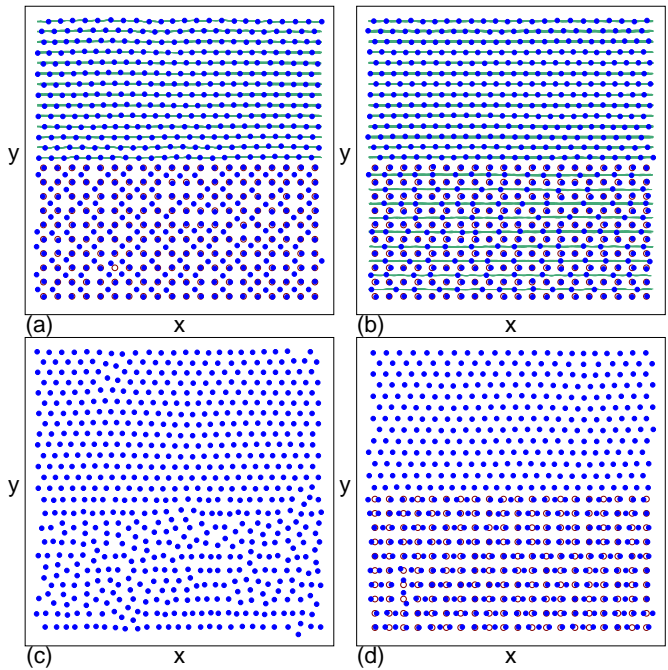}
\caption{
  The vortex positions (filled circles), pinning site positions (open circles),
  and vortex trajectories (lines) 
  for the system in Fig.~\ref{fig:1}(b) with parallel driving
  at $B/B_{\phi}  = 1.67$.
  (a) At $F_{D} = 0.1$, only the vortices in the unpinned region
  flow.
  (b) At $F_{D} = 0.5$, the vortices in the unpinned region are flowing along with the
  interstitial vortices in the pinned region.
  (c) A plot of only the vortex positions at
  $F_{D} = 1.2$, showing
  a coexistence between a moving hexagonal lattice
  in the upper, unpinned portion of the sample and
  a moving liquid or disordered state in the lower, pinned portion of the sample.
  (d) At $F_{D} = 1.2$ and $B/B_{\phi} = 1.33$,
  we find a moving lattice in the unpinned region and a
moving smectic in the pinned region.  
}
\label{fig:4}
\end{figure}

In Fig.~\ref{fig:4}(a) we show the motion in the sample from Fig.~\ref{fig:2}
at $B/B_{\phi} = 1.67$ and $F_{D} = 0.1$.
The vortices in the unpinned 
region are mobile and the vortices in the pinned region are immobile.
The pinned region is filled with
a combination of directly pinned 
vortices and interstitial vortices, whose motion is arrested only by interactions
with the neighboring pinned vortices.
In Fig.~\ref{fig:4}(b), the same sample at $F_{D} = 0.5$
contains moving interstitial vortices in the pinned region, but the vortices at the
pinning sites remain immobile.
When $F_{D} > 1.0$, all the vortices are moving; however,
we find that there is
a combination of a moving crystal state and a moving liquid state due to the
inhomogeneous pinning, as shown in Fig.~\ref{fig:4}(c) at $F_{D} = 1.2$.
Here
the vortices in the unpinned portion of the sample form a moving crystal
while the vortices in the pinned region are disordered 
and constantly exchange neighbors to form a moving liquid state.
A transverse pinning effect is responsible for the appearance of the moving
liquid.  Vortices in the pinned region preferentially
move along the easy-flow one-dimensional (1D)
channels separating neighboring rows of pinning sites; however, at this vortex
density, the spacing between the vortices is not commensurate with the spacing between
the easy-flow channels,
so the system remains disordered.
At higher drives than what we consider here, it is possible that the system could
order into a moving lattice in the unpinned portion of the sample
coexisting with a moving smectic state in the pinned portion of the sample.
The ability of the vortices
in the pinned region to dynamically order depends on the filling factor,
and certain fillings permit
ordered flows to occur in both the pinned and unpinned regions.
An example of this appears in
Fig.~\ref{fig:4}(d) for the same drive
of $F_D=1.2$ as in Fig.~\ref{fig:4}(c) but
at a field of $B/B_{\phi} = 1.33$, where the vortices in the pinned region can form
aligned 1D chains.

\begin{figure}
\includegraphics[width=3.5in]{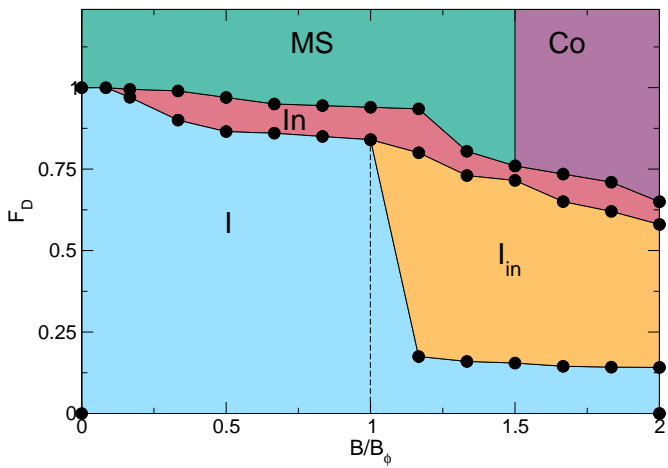}
\caption{
  A dynamic phase diagram as a function of $F_D$ vs $B/B_{\phi}$ for
  the system in Fig.~\ref{fig:2} with parallel driving.
  In Phase I (blue),
  vortices are only flowing in the pin-free region.
  The dashed line indicates the point at which interstitial vortices appear
throughout the pinned region in phase I
  when $B/B_{\phi} > 1.0$.
  In Phase I$_{in}$ (orange), both the vortices in the unpinned region and
  the interstitial vortices in the pinned region are flowing,
  but the vortices in the pinning sites remain pinned, as illustrated in
  Fig.~\ref{fig:4}(b).
  Phase In (pink) consists of 
  plastic or intermediate flow.
  In Phase MS (teal), all the vortices are moving and there is a combination of a
  moving triangular lattice in the unpinned region and a moving smectic in
  the pinned region, as shown in Fig.~\ref{fig:4}(d).
  Phase Co (purple) is a coexistence between a moving triangular lattice in
  the unpinned region and a moving liquid  in the pinned region, as illustrated
  in Fig.~\ref{fig:4}(c).
}
\label{fig:5}
\end{figure}

In Fig.~\ref{fig:5} we plot a dynamic phase diagram
as a function of $F_{D}$ versus $B/B_{\phi}$ for the system in Fig.~\ref{fig:2}
with parallel driving.
Here Phase I corresponds to the state in which only the vortices in the pinned
region are flowing,
as shown in Fig.~\ref{fig:3}(a) and Fig.~\ref{fig:4}(a).
The vertical dashed
line indicates the point at which immobile interstitial vortices appear throughout the
pinned region
for $B/B_{\phi} > 1.0$.
In Phase I$_{in}$, both the vortices in the unpinned region and the interstitial vortices
in the pinned region are flowing, while the pinned vortices remain immobile,
as shown in Fig.~\ref{fig:4}(b).   
Additional commensurate-incommensurate transitions of the interstitial vortex
structure can occur within phase I$_{in}$, producing small steps in the velocity-force
curves (not shown).
Phase In is an intermediate state in which different types of plastic flow occur.
The vortices can be disordered or partially disordered within the pinned region or at the
boundaries between the pinned and unpinned regions.
The distinct types of plastic flow states that occur within Phase In are not discussed
here but they have been outlined in previous work on samples that are filled with uniform
periodic pinning arrays \cite{Reichhardt97,Gutierrez09,Avci10,Reichhardt17}.
In Phase MS,
all of the vortices are moving and a triangular lattice in the unpinned region coexists
with a moving smectic phase in the pinned region,
as illustrated in Fig.~\ref{fig:3}(b) and Fig.~\ref{fig:4}(d).
In Phase Co, all the vortices are moving and there is a coexistence between a moving
triangular lattice in the unpinned region and a moving liquid in the pinned region.

\section{Driving in the Parallel Direction and Jammed States}

\begin{figure}
\includegraphics[width=3.5in]{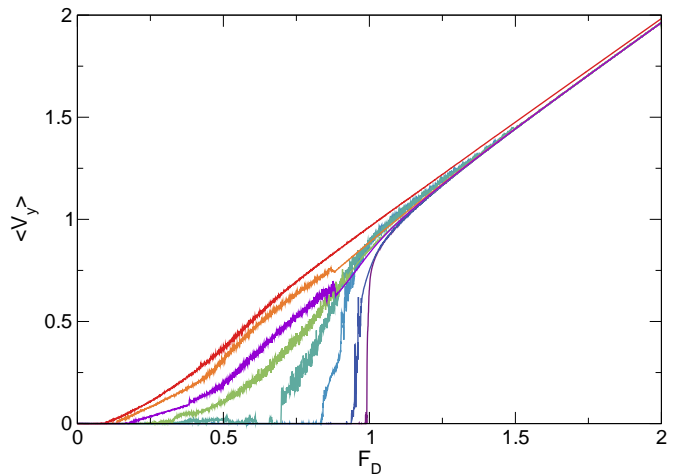}
\caption{
  $\langle V_{y}\rangle$ vs $F_{D}$ for the system in
  Fig.~\ref{fig:1} for driving in the perpendicular direction
  at $B/B_{\phi} = 0.083$, 0.167, 0.33, 0.5, 0.67, 0.83, 1.167,  and  $1.67$, from
  bottom to top.
  For $B/B_{\phi} < 0.5$, the system is pinned at low drives,
  while for $B/B_{\phi} > 0.5$, the system is jammed at low drives.  
}
\label{fig:6}
\end{figure}

We next consider driving in the positive $y$-direction, perpendicular to the
pinning strip, which causes
vortices in the unpinned region to
move toward the the pinned region.
In Fig.~\ref{fig:6} we plot $\langle V_y\rangle$ vs $F_{D}$
for $B/B_{\phi} = 0.083$, 0.167, 0.33, 0.5, 0.67, 0.83, 1.167, and $1.67$,
where, unlike the case of parallel driving, we find that there is a finite depinning
threshold for motion at all densities.
Figure~\ref{fig:6} indicates that the depinning
threshold decreases with increasing $B/B_{\phi}$.

Below the depinning threshold, we draw a distinction between pinned and jammed
phases.
The pinned phase appears when $B/B_{\phi} \leq 0.5$, and is characterized by the direct
pinning of vortices from the pin-free region by unoccupied pinning sites in the
pinned region.
At low drives, vortices in the unpinned region accumulate along the edge of the pinned
region and gradually force their way into the pinned region as the drive increases
above $F_D=0.33$.
As long as there are enough empty pinning sites to accommodate these vortices,
$\langle V_y\rangle = 0$.
At $B/B_{\phi} = 0.5$,
we find a combination of jammed and pinned states in which
a large portion of the vortices in the pin-free channel
accumulate along the edge of the pinned region and force
their way in, leading to a transient flow;
however, most of the vortices can eventually find unoccupied pinning sites.
These transients appear as a low level of flow near $F_D=0.5$ in 
the $B/B_{\phi} = 0.5$ curve in Fig.~\ref{fig:6} which is
followed at higher drives by a collapse of $\langle V_y\rangle$ back to zero,
indicating that all of the vortices have ceased to flow.
At $B/B_{\phi}> 0.55$, a clear jammed state emerges
when all the pinning sites become occupied
and the remaining vortices are immobilized only
due to interactions with vortices in the pinned region.
In systems that exhibit jamming,
flow is blocked by interactions with other particles rather
than being blocked by the direct trapping of particles by pinning sites 
\cite{Liu10,Reichhardt14}.
As the vortex density increases,
it becomes more difficult for the pinned vortices to hold back the vortices
in the unpinned region, and the
depinning threshold drops.
Jamming phases in vortex systems have been discussed previously
in the context of funnel geometries
\cite{Reichhardt10a,Karapetrov12,VlaskoVlasov13,Reichhardt18}
which contain constrictions that can block the vortex flow if the
vortex density in the constriction becomes too large.
Our results suggest that jamming can be observed clearly in
a system with inhomogeneous pinning, and that jamming occurs when
the number of vortices is larger than the number of pinning sites.

\begin{figure}
\includegraphics[width=3.5in]{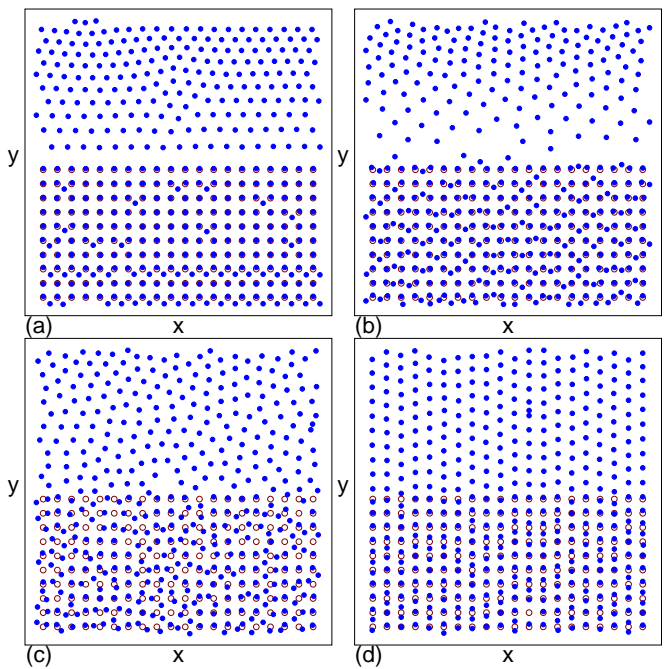}
\caption{
The vortex positions (filled circles) and pinning site positions (open circles)
for the system in Fig.~\ref{fig:6} with perpendicular driving  at $B/B_{\phi} = 1.167$.
(a) Jammed phase at $F_{D} = 0.125$. (b) Flowing phase at $F_{D} = 0.2$.
(c) Disordered uniform flow phase at $F_{D} = 0.75$.
(d) Moving flow or moving smectic phase at $F_{D} = 1.1$.
}
\label{fig:7}
\end{figure}

In Fig.~\ref{fig:7} we illustrate the vortex and pinning site locations for a
system with perpendicular driving at $B/B_{\phi} = 1.167$.
In Fig.~\ref{fig:7}(a), which shows the jammed state at $F_{D} = 0.125$,
vortices in the unpinned region pile up behind the vortices in the pinned region,
creating a vortex gradient in the unpinned region.
At the onset of motion for $F_D=0.2$,
Fig.~\ref{fig:7}(b) indicates
that there is still a gradient in the vortex density in the unpinned region,
but that vortices can move through the pinned region by forming interstitials
that slip between the pinning sites.
At $F_D=0.75$ in Fig.~\ref{fig:7}(c),
the flow is disordered and the gradient in the vortex density is lost.
Finally, at $F_D=1.1$ in
Fig.~\ref{fig:7}(d),
the vortices have dynamically ordered into a moving
smectic phase.

\begin{figure}
\includegraphics[width=3.5in]{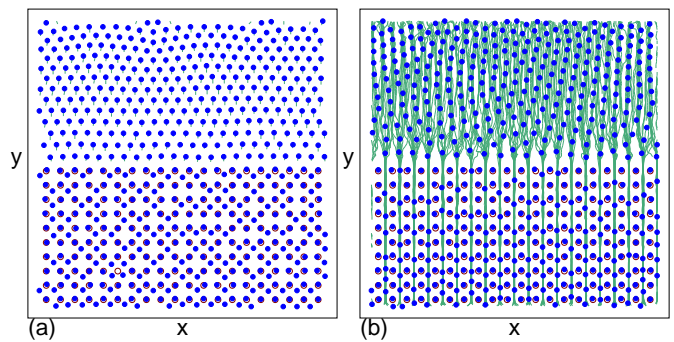}
\caption{ 
  The vortex positions (filled circles), pinning site positions (open circles),
  and vortex trajectories (lines) 
  in a system with perpendicular driving at $B/B_{\phi} = 1.67$. 
  (a) The jammed phase at $F_{D} = 0.1$.
  (b) The interstitial flow phase at $F_{D} = 0.2$.  
}
\label{fig:8}
\end{figure}

For $B/B_{\phi} > 1.33$ under perpendicular driving, the jammed phase depins into a
moving interstitial flow phase, where vortices can slip between the
pinned vortices,
and at higher drives the vortices at the pinning sites begin to depin, resulting in
disordered flow.
When $F_{D}/F_{p} >1.0$, the system can order into a moving smectic state.
In Fig.~\ref{fig:8}(a) we illustrate the jammed phase for a sample with perpendicular driving
at $B/B_{\phi} = 1.67$, where we trace out the vortex trajectories 
over a fixed period of time.
As the drive increases in the jammed phase,
the vortices in the unpinned region gradually become more compressed,
and in some cases they can undergo sudden avalanche-like rearrangements,
but the motion always returns to zero in the long time limit.
Figure~\ref{fig:8}(b) shows the occurrence of interstitial flow
at $F_{D} = 0.2$, where the vortices
follow 1D channels between the pinning sites.
As the vortices reenter
the pin-free region, the trajectories 
become more disordered and the vortices rearrange into a triangular lattice
within the unpinned region.

\begin{figure}
\includegraphics[width=3.5in]{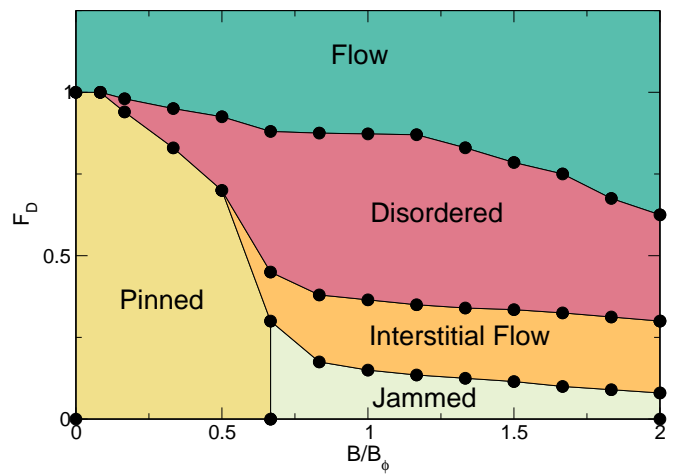}
\caption{ 
  A dynamic phase diagram as a function of $F_D$ vs $B/B_{\phi}$ for
  the system in Fig.~\ref{fig:6} with perpendicular driving.
  Labels indicate the
  pinned phase (yellow),
  jammed phase (pale green),
  interstitial flow phase (orange),
  disordered flow phase (pink),
  and free flow or moving smectic phase (teal). 
}
\label{fig:9}
\end{figure}

In Fig.~\ref{fig:9} we plot the dynamic phase diagram as a function of $F_D$
versus $B/B_{\phi}$ for the system in Fig.~\ref{fig:6} with perpendicular
driving.  Labels indicate the pinned
phase that appears for $B/B_{\phi} < 0.67$,
the jammed phase that occurs when $B/B_{\phi} \geq 0.67$,
a disordered or intermediate phase
composed of a mixture of pinned and flowing vortices,
the interstitial flow phase which
appears for drives just above the end of the jammed phase,
and the free flow or moving smectic state at high drives.

Several additional effects could be studied for the jamming system.
For example,
the vortex gradient in the unpinned region has the same features as the
conformal lattice that has been studied previously in vortex systems with
density gradients \cite{Ray13,Menezes17}.
Within the jammed phase, there could be avalanches with interesting
properties that occur when
the system tries to simultaneously
reduce the vortex lattice spacing
while maintaining a triangular vortex lattice configuration.

\section{Driving at an Angle}

\begin{figure}
\includegraphics[width=3.5in]{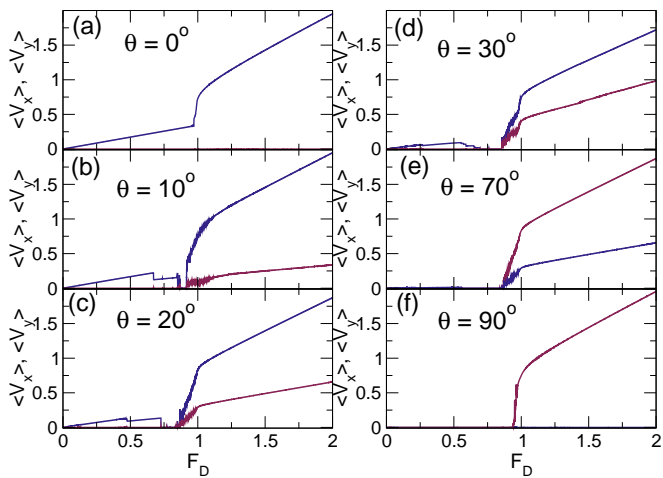}
\caption{
  $\langle V_{x}\rangle$ (blue) and $\langle V_{y}\rangle$ (red) vs $F_D$
  for the system in Fig.~\ref{fig:1} at $B/B_{\phi} = 0.167$ under driving
  at varied angle $\theta$ with respect to the $x$ axis.
  Here $\theta=$
  (a) $0^\circ$,
  (b) $10^\circ$,
  (c) $20^\circ$,
  (d) $30^\circ$,
  (e) $70^\circ$, and (f)  $90^\circ$.
  A reentrant pinning regime appears at intermediate values of $\theta$.  
}
\label{fig:10}
\end{figure}

We next consider the case where the drive is
applied at an angle $\theta$ with respect to the pinning strip.
In Fig.~\ref{fig:10} we focus on samples with
a low vortex density of $B/B_{\phi} = 0.167$.
We plot $\langle V_x\rangle$ and $\langle V_y\rangle$ versus $F_D$ in
Fig.~\ref{fig:10}(a)
for $\theta = 0^\circ$, where the two low density-dynamic phases I and MS
described above in section III appear.
At $\theta = 10^\circ$, 
Fig.~\ref{fig:10}(b) indicates that the vortices are guided along the
$x$ direction when they first begin to move,
but as $F_D$ increases, there is a downward step in
$\langle V_x\rangle$ followed by a drive-induced pinned region
where both $\langle V_{x}\rangle$ and $\langle V_{y}\rangle$ are zero.
Here, when $F_D$ is small,
the vortices in the unpinned region undergo easy flow in the $x$-direction
that is guided by the repulsion from the vortices trapped in the pinned region.
As $F_D$ increases, the
flowing vortices are eventually pushed into the pinned region,
where they can each find an empty pinning site and become 
immobile, producing a drop in the vortex mobility.
Since there 
are many more pinning sites than vortices,
it is possible for all of the vortices to become trapped by pinning sites,
producing the reentrant pinned phase.
When the drive is high enough,
the vortices at the pinning sites become depinned and both
$\langle V_x\rangle$ and $\langle V_y\rangle$ begin to increase with
increasing $F_D$.
Similar behavior appears in
Fig.~\ref{fig:10}(c,d) 
for $\theta = 20^\circ$ and $\theta=30^\circ$,
with the width of the reentrant pinning regime increasing as $\theta$ increases.
At $\theta=70^\circ$ in Fig.~\ref{fig:10}(e),
the guided flow phase is almost completely lost, and  at  
$\theta = 90^\circ$ in Fig.~\ref{fig:10}(f), there is only a single pinned phase and
the reentrance disappears.

\begin{figure}
\includegraphics[width=3.5in]{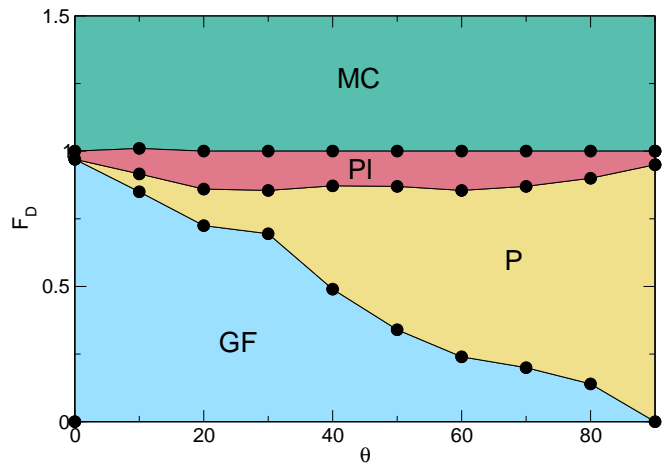}
\caption{
  A dynamic phase diagram as a function of $F_D$ vs $\theta$ for
  the system in Fig.~\ref{fig:10} with $B/B_\phi=0.167$.
  Labels indicate
  the guided flow phase GF (blue),
  the pinned phase P (yellow),
  the intermediate or plastic flow phase Pl (pink),
  and the free flow phase MC (teal).
}
\label{fig:11}
\end{figure}

In Fig.~\ref{fig:11} we plot the dynamic phase diagram as a function
of $F_D$ versus $\theta$
for the system in Fig.~\ref{fig:10} with $B/B_\phi=0.167$.
We find a guided flow phase, a pinned phase,
a moving crystal or free flow phase at higher drives,
and an intermediate or 
plastic phase between the pinned and free flow phases.
The width of the
reentrant pinned phase increases
with increasing $\theta$.

\begin{figure}
\includegraphics[width=3.5in]{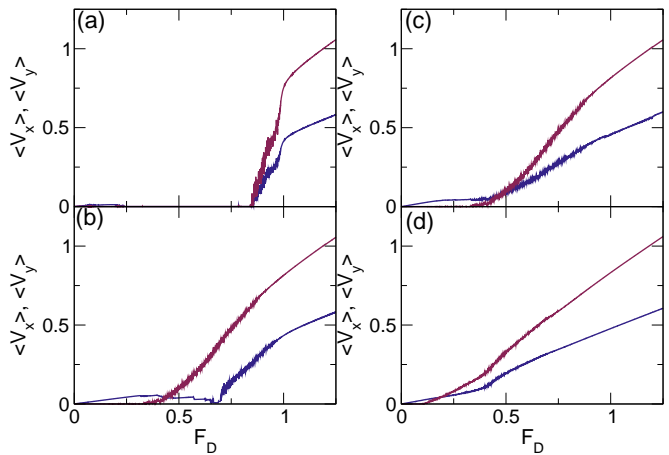}
\caption{
  $\langle V_{x}\rangle$ (blue) and $\langle V_{y}\rangle$ (red) vs $F_D$
  for driving at $\theta=60^\circ$ with
  $B/B_{\phi} =$
  (a) 0.167,
  (b) $0.33$,
  (c) $0.67$, and (d) $1.67$.
}
\label{fig:12}
\end{figure}

The behavior of the reentrant pinning phase also depends on the filling factor. 
In Fig.~\ref{fig:12}(a) we plot $\langle V_{x}\rangle$
and $\langle V_{y}\rangle$ versus $F_{D}$ for the system in Fig.~\ref{fig:10} at 
$B/B_{\phi} = 0.167$ and
$\theta = 60^\circ$, which is in a pinned state at most low drives.
When $B/B_{\phi}$ is increased to
$B/B_{\phi} = 0.33$, Fig.~\ref{fig:12}(b) indicates that an
extended region emerges in which
guided flow in the $x$-direction occurs.
This is followed by a drive window over which
$\langle V_{x}\rangle$ decreases with increasing $F_D$, indicating
the appearance of
negative differential conductivity, while
$\langle V_y\rangle$ increases with increasing $F_D$.
For 
$F_{D} > 0.7$, both $\langle V_{x}\rangle$ and $\langle V_{y}\rangle$
increase with increasing $F_D$.
For $F_{D} > 1.0$, the velocity increase becomes linear with $F_D$ as the
system enters
the free flow limit.
At $B/B_\phi=0.67$ in
Fig.~\ref{fig:12}(c),
there is still a guided flow state at low drives,
but the window of negative differential mobility is reduced in width.
In Fig.~\ref{fig:12}(d)
at $B/B_{\phi} = 1.67$,
guided flow appears for
low $F_{D}$  where $\langle V_{y}\rangle = 0$,
and at higher drives, both velocity components increase monotonically with
increasing $F_D$.
We note that for certain
driving angles, we also observe directional locking effects.
We do not discuss these effects here, but similar features
have been 
analyzed in previous woks \cite{Reichhardt99,Reichhardt08a}. 

\section{Summary}
We have numerically examined the dynamics of vortices in a system containing
a square array of pinning sites where a strip of the pinning sites are
removed in order to create an easy flow channel for vortex motion.
When a driving force is applied parallel to the pinning strip,
flow first occurs for
the vortices in the unpinned region,
followed by the depinning
of vortices in the pinned region, 
leading to velocity-force curves with a  two step feature.
At high vortex densities we find additional phases including
a multiple step depinning transition within the pinned region. 
For driving applied perpendicular to the pinning strip,
we find that either a pinned or jammed state occurs at low drives depending on the
vortex density.
The jammed phase appears at higher magnetic fields
when vortices in the unpinned region are blocked by the pinned vortices and
accumulate at the edge of the pinned region,
while at lower fields, a pinned state appears in which all of the vortices are
eventually trapped directly by pinning sites.
The crossover from the pinned to the jammed state appears as a
decrease in the critical depinning force and an increase in the nonlinearity of
the velocity-force curves.
When the drive is applied at an angle to the pinning strip, we find
a dynamically induced reentrant pinning,
where vortices undergo guided flow at low drives due to the repulsion from
the pinned vortices,
but gradually enter the pinned region and become trapped as the driving
force increases.
If the number of vortices is smaller than the number of pinning sites, all of the
vortices can become trapped, producing the reentrant pinned state.
At higher vortex densities,
the reentrant pinning effect is gradually replaced by
a region of negative differential mobility and then by monotonic
anisotropic flow.
Our results could be tested using nanostuctured pinning arrays
in which a portion of the pinning sites are omitted.

\acknowledgments
This work was supported by the US Department of Energy through
the Los Alamos National Laboratory.  Los Alamos National Laboratory is
operated by Triad National Security, LLC, for the National Nuclear Security
Administration of the U. S. Department of Energy (Contract No. 892333218NCA000001).

\bibliography{mybib}

\end{document}